\documentclass[10pt]{article}
\usepackage{amsmath}
\usepackage{algorithm}
\usepackage{algpseudocode}
\usepackage{xcolor}
\usepackage{amsmath} 
\usepackage{array}
\usepackage{caption}
\usepackage{booktabs}
\usepackage{multirow}
\usepackage{graphicx}
\usepackage{threeparttable}
\usepackage[inkscapelatex=false]{svg}
\usepackage[letterpaper]{geometry}
\usepackage{hicss}
\usepackage{times}
\usepackage[none]{hyphenat}
\usepackage{url}
\usepackage{latexsym}
\usepackage{minted}
\usepackage{indentfirst}
\usepackage{graphicx}
\graphicspath{{images/}}
\usepackage[
    style=apa,
  ]{biblatex}
\title{Collaborative LLM Agents for C4 Software Architecture Design Automation}

\author{Kamil Szczepanik \\
 Warsaw University of Technology \\
 {\underline{ kamil.szczepanik.stud@pw.edu.pl}} \\ \And
 Jarosław A. Chudziak \\
Warsaw University of Technology \\
 {\underline{ jaroslaw.chudziak@pw.edu.pl} } \\ 
}

\date{}

\begin{document}
\maketitle
\begin{abstract}

Software architecture design is a fundamental part of creating every software system. Despite its importance, producing a C4 software architecture model~--- the preferred notation for such architecture --- remains manual and time-consuming. We introduce an LLM-based multi-agent system that automates this task by simulating a dialogue between role-specific experts who analyze requirements and generate the Context, Container, and Component views of the C4 model. Quality is assessed with a hybrid evaluation framework: deterministic checks for structural and syntactic integrity and C4 rule consistency, plus semantic and qualitative scoring via an LLM-as-a-Judge approach. Tested on five canonical system briefs, the workflow demonstrates fast C4 model creation, sustains high compilation success, and delivers semantic fidelity. A comparison of four state-of-the-art LLMs shows different strengths relevant to architectural design. This study contributes to automated software architecture design and its evaluation methods.

\end{abstract}

\subsubsection*{Keywords:}

LLM agents, Multi-Agent Systems, Software Architecture Design, C4 Model

\section{Introduction}

Designing software architecture is a fundamental process at the start of every software system.
Software Architecture Design (SAD) is a stage of software creation when all the main components, stakeholders, and all the elements for fulfilling functional and non-functional requirements are addressed (\cite{bass2021software}).  This is when engineers define interactions between components, the technology stack, and how it could achieve scalability, maintainability, and security.  Methodologies and frameworks have been developed to make this process more structured and controlled, e.g., the C4~model (\cite{brown2018c4}), TOGAF (\cite{josey2016togaf}), and the 4+1 view model (\cite{kruchten20024+}).
The C4~model is a framework for visualizing and communicating the architecture at different levels of abstraction --- Context, Container, Component, Code.  By well-made decomposition of the system into abstraction levels, it is possible to parallelize work and narrow its scope.  The C4 model has been and still is one of the most important tools in architecting software.

For experienced engineers, this task may not be particularly hard, but it is still effortful when designing complex systems. Usually, it requires professionals with different expertise. Moreover, different levels of abstraction often demand different areas of expertise (\cite{pargaonkar2023enhancing}); e.g., at the Context level, a business perspective is required to correctly define all stakeholders of the system. Although the C4 model was made to control the design process, the complexity of systems can still lead to inconsistencies and errors in definition. The development of C4 levels is relatively slow, especially when drafting the first version of a complex system. Automation of the process could significantly help engineers and accelerate the work. These challenges motivate this research.
By providing a first draft of a C4 model at an early stage, developers could simply validate or correct the diagrams according to their knowledge. An evaluation of the draft would be valuable as well, because it could tell the users about the flaws and quality of the model they received. This could allow skilled engineers to focus on higher-level critique and refinement rather than routine diagramming.

The recent advances in Large Language Models (LLMs) show their substantial potential in analytical, creative, and routine tasks. To tackle more complex tasks, researchers study LLM-based multi-agent systems (MAS) in various domains such as problem solving and world or environment simulation. This includes studies on LLM agents following software engineering workflows (\cite{cinkusz2024cognitive}), performing creative and analytical tasks (\cite{wang2024survey, szczepanik2025triz}), and, on the other hand, simulating realistic scenarios in social sciences or human economic behavior (\cite{park2023generativeagentsinteractivesimulacra, DeFortuny2025Simulating}). It is crucial to understand that with the improving capabilities of LLMs, the tasks of future software developers and engineers will change. The human-AI collaboration tools will work as a productivity enhancement rather than a full replacement for human experts. In AI-assisted development, a programmer working with an AI collaborator can finish tasks in half the time of someone working alone (\cite{peng2023impact}).

In this research, we propose a C4 model generation and evaluation method. Our LLM-powered multi-agent system simulates the requirement analysis and diagram creation. A system brief, with a description and functional and non-functional requirements, is the input to the system. For three out of four C4 abstraction levels, we simulate a conversational analysis of the element at stake, with a number of specialized agents. Agents simulate realistic personas, which gives multiple perspectives for analysis. The output of the system is a set of artifacts for each abstraction level --- textual analysis, YAML structure, and PlantUML diagram describing the element at stake. As the second contribution, our research provides an evaluation method, including deterministic checks for structural and syntactic integrity and C4 rule consistency, and semantic and qualitative scoring via an LLM-as-a-Judge approach (\cite{zheng2023judging}). Automated evaluation of such data is problematic because it requires assessing more than just structural or syntactic correctness. It must also judge semantic meaning, conceptual plausibility, and qualitative clarity, which are not easily measured by traditional deterministic checks. 
Our research questions (RQs) are as follows. \textbf{RQ1:} To what extent can a multi-agent system generate a C4 model, and how do LLM models influence the quality? \textbf{RQ2}: Can a multi-agent LLM approach outperform a single-agent baseline in C4 model generation? 
Our work also contributes by providing prompt engineering guidelines and a comparative study of several LLMs, identifying the models' advantages and disadvantages.

The remainder of the paper is organized as follows. Section 2 reviews background and related work. Section 3 details the proposed C4 generation workflow and evaluation method. Section 4 describes the experimental setup. Section 5 presents the results, and Section 6 discusses the outcomes and future work. Finally, Section 7 concludes our research.

\section{Related Work}

In this section, we introduce the background of the C4 model and software architecture design, and survey prior literature on applications of LLMs and LLM-based multi-agent systems in software engineering.

\begin{figure*}[htbp]
  \centering
  \includegraphics[width=\linewidth]{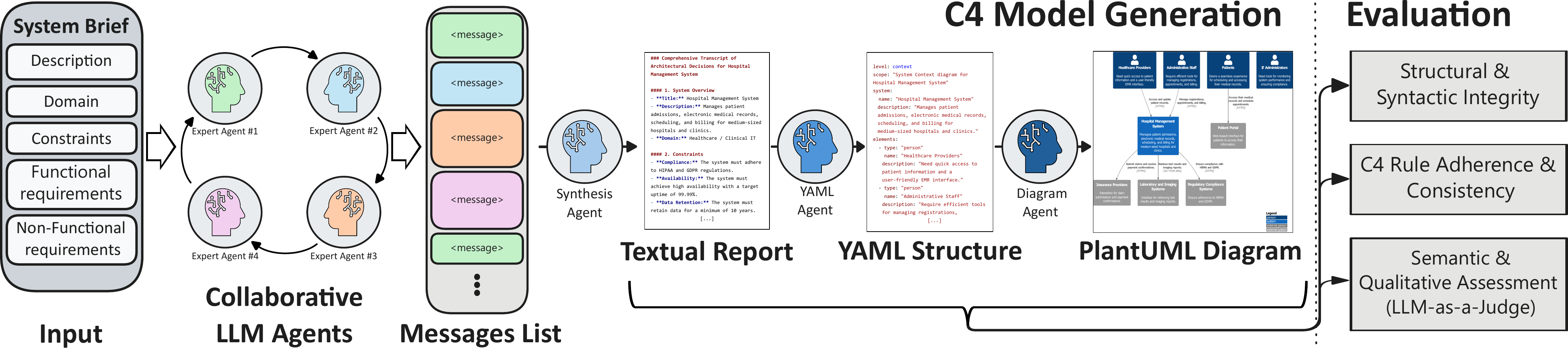}
  \caption{The proposed system workflow, illustrating the multi-agent C4 model generation pipeline and the subsequent evaluation framework. The generation stage (left) shows expert agents collaboratively processing a system brief, with specialized agents producing C4 artifacts and diagrams, while the evaluation stage (right) assesses these artifacts using structural, rule-based, and semantic criteria.}
  \label{fig:diagram_overview}
\end{figure*}

The C4 model makes software architecture design more structured by abstracting a system into four hierarchical levels --- Context, Containers, Components, and Code. 
The C4 model also improves communication among stakeholders and has become a standard approach for managing the complexity of modern software systems (\cite{brown2018c4}). On the Context level (L1), designers identify the system’s external actors --- people and software systems that interact with it --- and articulate the core goals and usage scenarios that frame those interactions. On the Container level (L2), they map the major runtime executables and data stores --- web services, databases, background workers, message brokers --- annotating each container’s responsibilities, chosen technology, and communication paths. The Component level (L3) breaks each container into cohesive components, detailing their responsibilities, interfaces, and interaction patterns to ensure modularity and separation of concerns. Lastly, the Code level (L4) zooms into implementation artifacts --- classes, modules, functions, and configuration files --- revealing how each component is realized in code while retaining traceability to the higher architectural views.

Previous research has shown that software engineering methodologies originally designed for humans can be incorporated into agentic LLM workflows (\cite{cinkusz2024cognitive}). With the rapid advancement of Large Language Models, the agentic approach has proven especially successful: recent work reports that agentic frameworks outperform state-of-the-art baselines across several domains, including software development (\cite{qian2023communicative}). Other sources argue that agentic workflows can achieve next-generation model performance (\cite{wu2023autogen}). Previous studies on reasoning and problem-solving with LLMs document how prompt strategies, such as Chain-of-Thought or ReAct (\cite{wei2022chain, yao2022react}), significantly shape their outputs.

Similarly, research demonstrates that generative AI boosts productivity in intermediate-level writing and coding tasks (\cite{dell2023navigating}). Collaborative AI has also been shown to boost organisational effectiveness, for example, within workplace environments where it enhances automation and decision support (\cite{Svensson2024Agentic}). These systems excel at analysing large volumes of data and generating solutions, and extensive reviews describe practical patterns of AI–human collaboration (\cite{davenport2022working}).
The use of LLMs for automated evaluation, the LLM-as-a-Judge approach (\cite{zheng2023judging}), is a rapidly expanding research area, with applications ranging from automated judgements to entity extraction and relation identification (\cite{gu2024survey}).
Human–AI collaboration tools are a valuable support, but without human supervision, they remain limited. Replacing the human partner with another AI agent yields a multi-agent system (MAS) --- a computer system capable of tackling more complex tasks. A key strength of an LLM-based MAS is conversational collaboration: agents interact, debate, and exchange ideas, thereby simulating the work of a team (\cite{hong2023metagpt}).

To our knowledge, no previous work has presented an LLM-based multi-agent system for automated C4 model generation from a system brief as input, nor has it presented an evaluation method for its outputs. Research on the evaluation of C4 models is scarce, while in the age of automation, it will be necessary. This research contributes to filling this gap.

\section{Methodology}

In this section, we present our approach --- an LLM-based multi-agent system for C4 model generation --- and the method we use to evaluate its outputs. First, we define the system and identify its crucial components. Next, we describe the generation workflow. Finally, we introduce the evaluation protocol. To keep the study’s scope firmly on architectural reasoning, we focus on the first three levels of the C4 model. Generating code-level (L4) diagrams would require a larger experimental platform and additional evaluation metrics beyond our current framework.

\subsection{Multi-Agent LLM Architecture}

Our system ($MAS_{C4}$) is modelled as a stateful system defined by the 3-tuple:
\begin{equation}
    MAS_{C4} = (L, A, S),
\end{equation}
where \(L\) is the set of C4 abstraction \emph{Levels} (\(L = \{L_{1}, L_{2}, L_{3}\}\)); \(A\) is the set of functional \emph{Agents}; and \(S\) is the set of all possible system \emph{States}. Each state \(s \in S\) is a self-contained snapshot that includes the complete history of all generated artifacts, representing the full context available for the next operation.

\subsubsection{System State \(\boldsymbol{S}\)}

A \emph{state} \(s \in S\) is a 4-tuple \(s = (M, b, \mathcal{A}, Q_c)\) that represents the complete operational context. The components are: (1) the \textbf{Message History} \(M\); (2) the \textbf{System Brief} \(b\); (3) the \textbf{Artifact Set} \(\mathcal{A}\), representing the system's complete and monotonically growing historical output; and (4) the \textbf{Component Queue} \(Q_c\), which holds container names from \(L_2\) to orchestrate \(L_3\) generation.

The initial state \(s_0\) is configured with an empty message history (\(M=\emptyset\)), an empty artifact set (\(\mathcal{A}=\emptyset\)), an empty component queue (\(Q_c=\emptyset\)), and the user-provided system brief \(b\). The workflow proceeds via a sequence of states \((s_0, s_1, \dots)\). An operation on state \(s_k\) produces the subsequent state \(s_{k+1}\) by appending a new message to \(M\) or a new artifact to \(\mathcal{A}\).

\subsubsection{Artifact Set \(\boldsymbol{\mathcal{A}}\)}

The \emph{Artifact Set} \(\mathcal{A}\), a component of the state, is append-only. Each artifact \(\alpha \in \mathcal{A}\) has one of the types:
\texttt{TRANSCRIPT}, \texttt{ANALYSIS\_REPORT}, \texttt{VIEW\_YAML}, or \texttt{PLANTUML\_DIAGRAM}.

\subsubsection{Agent Set \(\boldsymbol{A}\)}

An \emph{agent} \(a_{\pi,\tau} \in A\) is parameterized by persona \(\pi\) and task \(\tau\). With prompt templating \(P_T\) and a context map \(C\), the agent generates an output \(y\) based on the current state:
\begin{equation}
    y \;=\; a_{\pi,\tau}(s)
    \;=\; \mathcal{LLM}\!\Bigl(P_{T}\bigl(\pi,\,\tau,\,C(s)\bigr)\Bigr)
\end{equation}
where \(y\) is the task-dependent output (a message $m$ or an artifact $\alpha$). The $\mathcal{LLM}$ function represents a call to a large language model, which takes the fully assembled prompt from $P_T$ and returns a textual completion. The prompt templating function $P_T$ assembles the final prompt by combining the persona $\pi$, the task $\tau$, and the contextual data provided by $C(s)$. The context map \(C(s)\) selects a primary input (e.g., \(M\) for collaborative agents) and includes \(b\), the current level \(\ell\in L\), and any relevant prior artifacts selected from the state's complete artifact set \(\mathcal{A}\). For instance, when executing the \texttt{STRUCTURE\_YAML} task, $C(s)$ selects the relevant \texttt{ANALYSIS\_REPORT} from \(\mathcal{A}\), along with the \texttt{VIEW\_YAML} from the preceding C4 level to ensure structural consistency. After an agent generates its output, the system transitions to a new state by appending new messages to \(M\) or new artifacts to \(\mathcal{A}\). 

We partition the agent set \(A\) into two distinct archetypes that correspond to the two primary stages of the generation process: \emph{Collaborative Analysis Agents} (\(A_{collab}\)) and \emph{Specialized Processing Agents} (\(A_{proc}\)). First, collaborative agents perform the exploratory work of architectural design through simulated dialogue. Second, processing agents execute a deterministic transformation chain of artifact types to convert the dialogue's output into the final, structured artifacts.

\subsubsection{Collaborative Analysis Agents (\(A_{collab}\))}

Those agents simulate a human design workshop, performing the \texttt{ANALYZE} task to generate a transcript.

At each level $\ell \in L$ (and for each $L_3$ container from $Q_c$), a new collaborative session begins. The message list is reset to $M=\emptyset$, while the full artifact history \(\mathcal{A}\) from the prior state provides context via $C(s)$. Intuitively, this resembles a new expert meeting, where $M$ captures the live discussion and \(\mathcal{A}\) supplies the complete record of notes from all previous sessions.

Team personas are varied by level to provide tailored expertise. The L1 (System Context) team consists of a \texttt{Product Owner}, \texttt{Business Analyst}, and \texttt{Lead Software Architect}. The L2 (Containers) team includes a \texttt{Software Architect}, \texttt{Lead Developer}, \texttt{DevOps Specialist}, and \texttt{Security Specialist}. Finally, the L3 (Components) team is composed of a \texttt{Lead Developer}, \texttt{Senior Developer}, \texttt{Database Administrator}, and \texttt{Security Specialist}.

Let the task be $\tau=\texttt{ANALYZE}$. The team comprises $K$ agents
$a_{\pi_0,\tau}, a_{\pi_1,\tau}, \dots, a_{\pi_{K-1},\tau}$,
where agent $a_{\pi_i,\tau}$ has a fixed persona $\pi_i$.
The team dialogues for $N$ rounds in round-robin order. At the global turn
$k\in\{0,\dots,NK-1\}$, let $j=k \bmod K$ be the scheduled index. The scheduled agent generates the next message based on the current state of the dialogue:
\begin{equation}
\label{eq:collab_agent_corrected}
m_{k+1} = a_{\pi_j,\tau}(s_k)
= \mathcal{LLM}\!\Bigl(P_{T}\bigl(\pi_j,\,\tau,\,C(s_k)\bigr)\Bigr)
\end{equation}
The system then transitions to state $s_{k+1}$ by appending $m_{k+1}$ to the message history $M$. After all $N$ rounds
($NK$ messages), the final list $M$ constitutes the session’s \texttt{TRANSCRIPT}, which is then added as an artifact to \(\mathcal{A}\).

\subsubsection{Specialized Processing Agents ($A_{proc}$)}

These agents perform a single-pass transformation chain per level instance. Let the ordered sequence of processing tasks be $(\tau_1, \tau_2, \tau_3) =$ $(\texttt{SYNTHESIZE},$ $\texttt{STRUCTURE\_YAML},$ $\texttt{GENERATE\_PLANTUML})$. Each task $\tau_i$ in the sequence is performed by a dedicated agent with a tailored persona $\pi_i$ to simulate a real-world specialist $(\pi_1, \pi_2, \pi_3) =$ $(\texttt{Technical Writer},$ $\texttt{Software Architect},$ $\texttt{PlantUML Diagram Specialist})$

The $A_{proc}$ agent chain begins with the state $s_0$ resulting from the collaborative phase, thus containing the \texttt{TRANSCRIPT} artifact $\alpha_0$  ---  the primary input for the chain. The process unfolds sequentially for each step $i \in \{1, 2, 3\}$, where each agent builds upon the last one's work. At each step, the corresponding agent generates a new artifact:
\begin{equation}
\label{eq:proc_agent}
\alpha_{i} = a_{\pi_i,\tau_i}\!(s_{i-1})
= \mathcal{LLM}\!\bigl(P_{T}\bigl(\pi_i,\,\tau_i,\,C(s_{i-1})\bigr)\bigr)
\end{equation}
The context map $C(s_{i-1})$ provides the artifact $\alpha_{i-1}$ generated in the previous step as the primary input (for $i>1$). The system then transitions to the next state $s_i$, by appending the new artifact $\alpha_{i}$ to the state's artifact set $\mathcal{A}$. This sequential state update ensures that the full context is passed down the chain.

\begin{figure}[t]
    \centering
    \includegraphics[width=0.89\linewidth]{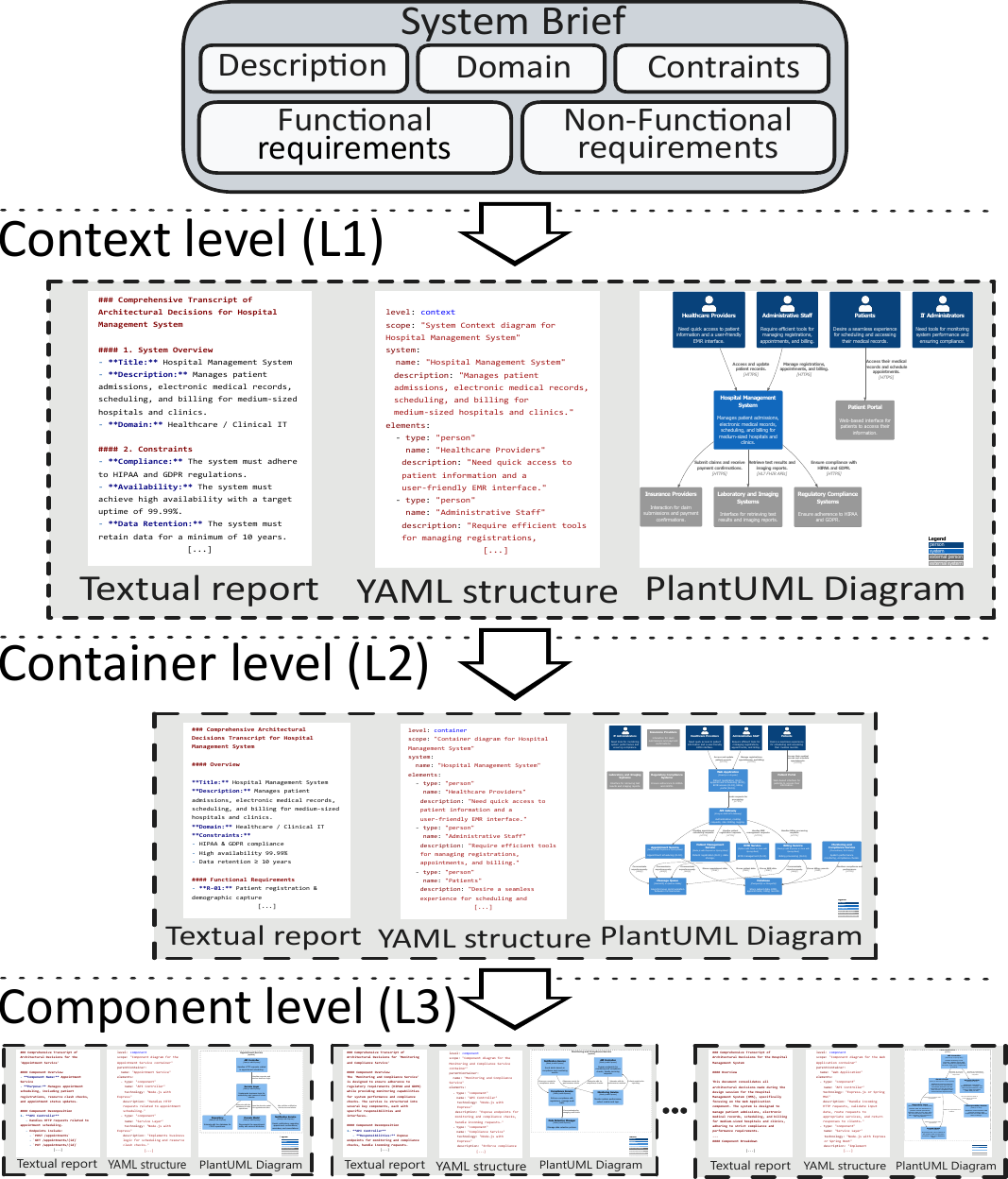}
    \caption{Top-down multi-level decomposition workflow for C4 architecture modeling.}
    \label{fig:c4_workflow}
\end{figure}

\subsection{C4 Model Generation Workflow}
\label{sec:workflow}

The system's workflow proceeds top-down through the C4 levels, as illustrated in Figure~\ref{fig:c4_workflow}. First, the core generation pipeline is executed for Level 1 (Context), using the initial brief \(b\) to produce L1 artifacts. Next, the pipeline runs for Level 2 (Containers), with the L1 artifacts providing context to ensure consistency. Finally, for Level 3 (Components), the system populates the \texttt{Component Queue} ($Q_c$) with all containers defined at L2 and executes the pipeline iteratively for each one, using the relevant L1 and L2 artifacts as context for each run.

Each pipeline execution involves an artifact generation chain (Figure \ref{fig:diagram_overview}). It begins with collaborative analysis, where a team \textit{Collaborative Analysis Agents} ($A_{collab}$) generates a \texttt{TRANSCRIPT} as defined in Eq.~\eqref{eq:collab_agent_corrected}. This is followed by a three-step processing chain where \textit{Specialized Processing Agents} ($A_{proc}$) perform transformations according to Eq.~\eqref{eq:proc_agent}: an agent executing the \texttt{SYNTHESIZE} task creates an \texttt{ANALYSIS\_REPORT}, a second performing the \texttt{STRUCTURE\_YAML} task produces a \texttt{VIEW\_YAML}, and a third on task \texttt{GENERATE\_PLANTUML} renders the final \texttt{PLANTUML\_DIAGRAM}. This structured workflow ensures that every generated artifact is traceable to a specific architectural level and context, resulting in a comprehensive C4 model of the target system.

\subsection{Prompt Engineering}
\label{sec:prompt_engineering}

The behavior and output quality of the agents in the $MAS_{C4}$ system are critically dependent on prompt engineering. Rather than relying on simple, single-sentence instructions, each prompt is a composite document constructed by the templating function $P_T$. This function assembles several key components to provide the LLM with the optimal context for its task.

The core of every agent prompt includes a \textbf{Persona} ($\pi$), a detailed narrative that instructs the LLM to adopt a specific expert role (e.g., \texttt{Software Architect}). This persona defines the agent's goals, focus areas, and contribution style. Next, a clear \textbf{Task} command ($\tau$) defines the agent's objective, such as \texttt{ANALYZE} or \texttt{STRUCTURE\_YAML}. Finally, the prompt's \textbf{Context} provides the situational information necessary for the task, assembled by the context map $C(s)$. This includes the message history ($M$) for collaborative agents, an input artifact ($\alpha_{in} \in \mathcal{A}$) for processing agents, and other relevant details from the state $s$, like the current C4 level.

Our methodology is built on a combination of established prompting strategies (\cite{white2023prompt}). The primary strategy is the use of detailed personas, as described above. For the processing agents responsible for artifact transformation, we employ a schema-guided zero-shot prompting strategy. Rather than providing solved input-output examples (a few-shot approach), the prompt gives the model a detailed set of instructions and a structural template or syntax guide, constraining the output to the desired format. Furthermore, the artifact generation chain can be viewed as a form of macro-level Chain-of-Thought (CoT) reasoning (\cite{wei2022chain}), where the output of one agent (e.g., the \texttt{ANALYSIS\_REPORT}) is a “thought” that becomes the direct input for the next agent in the sequence.

\subsection{Evaluation Module}
\label{sec:evaluation_module}

We propose a hybrid evaluation framework to assess the quality of AI-generated software architecture diagrams. It provides an overall view of performance by combining objective, automatable checks with qualitative and descriptive assessments using an LLM-as-a-Judge approach. Three layers are included, each assessing a different set of qualities: \textit{Structural \& Syntactic Integrity}, \textit{C4 Rule Adherence \& Consistency}, and \textit{Semantic \& Qualitative Assessment}.

The first layer, \textit{Structural \& Syntactic Integrity}, checks for basic correctness and completeness of the generated artifacts. \textit{Compilation Success} is evaluated by invoking the official PlantUML command-line runner for each diagram. A diagram is marked valid if the process completes successfully, and a per-diagram record is stored. \textit{C4 Model Completeness} verifies that, for each level, the expected artifacts --- \texttt{ANALYSIS\_REPORT}, \texttt{VIEW\_YAML}, and \texttt{PLANTUML\_DIAGRAM} --- are generated and non-empty. This is implemented as programmatic checks for the presence and non-emptiness of the expected artifacts within the resulting artifact set.

\begin{figure}[b!]
    \centering
    \includegraphics[width=0.97\linewidth]{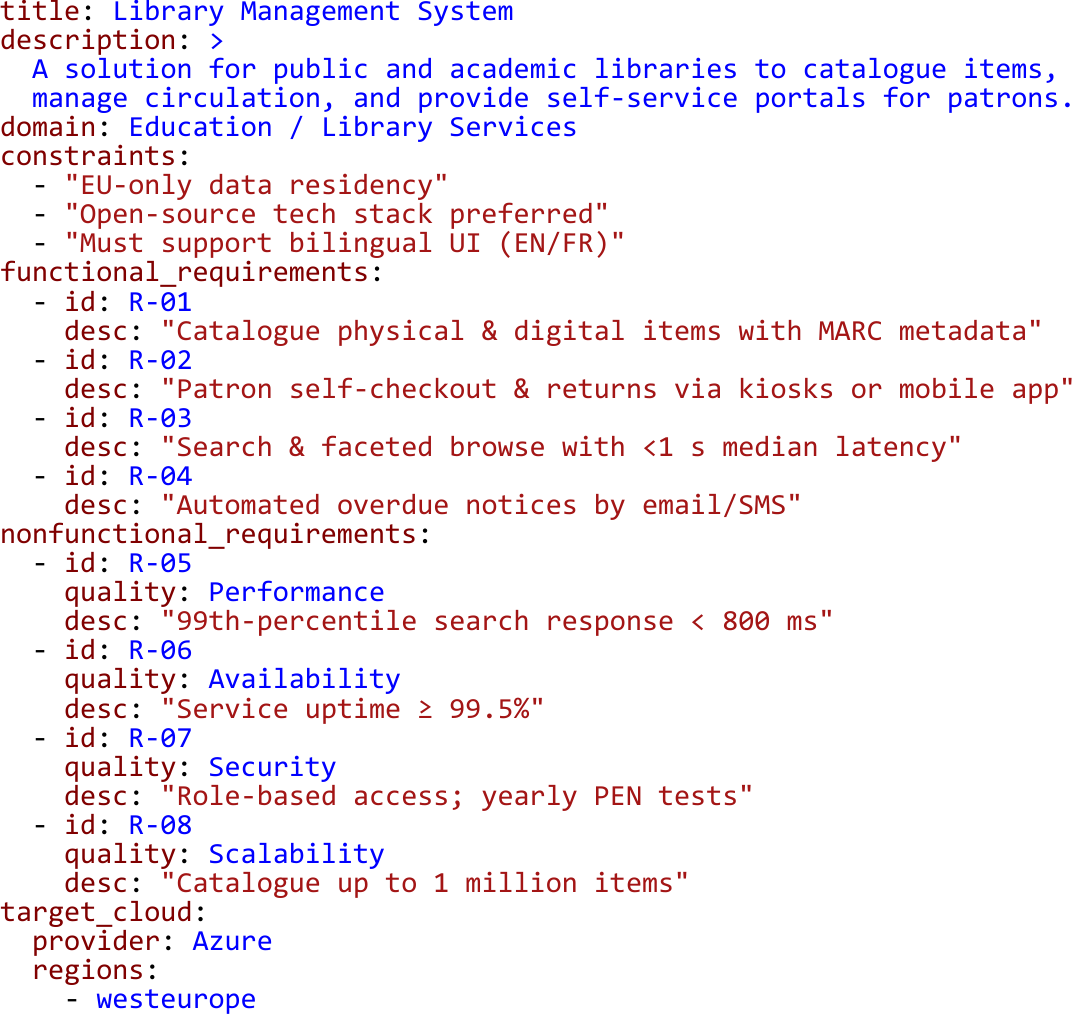}
	\caption{System brief input of the Library Management System test case.}
	\label{fig: input}       
\end{figure}

In the second \textit{C4 Rule Adherence \& Consistency} layer, generated C4 models are assessed to determine whether they adhere to the principles of the C4 methodology and maintain internal consistency. \textit{Abstraction Adherence} is implemented as rule-based regex checks over PlantUML source that flag level-inappropriate elements (e.g., \texttt{Component} in Context/Container) and required PlantUML constructs (e.g., \texttt{System\_Boundary}, \texttt{Container\_Boundary}). \textit{Emergent Naming Consistency} computes the dominant naming convention across element names (patterns for PascalCase, camelCase, snake\_case, kebab-case) and automatically flags outliers. \textit{Definitional Consistency} verifies that all elements defined in the \texttt{VIEW\_YAML} are present in its corresponding \texttt{PLANTUML\_DIAGRAM}. \textit{Cross-Level Consistency} verifies that externals shown at L2 (Container) exactly match those declared at L1 (Context), and that each L3 Component only includes elements defined at L2 (containers, people, external systems, databases).

The final layer, \textit{Semantic \& Qualitative Assessment}, uses an LLM-as-a-Judge approach to perform qualitative assessments that simulate human expert judgment. This layer includes \textit{Semantic Consistency}. It determines if the Level 1 Context diagram accurately captures the key entities mentioned in the original system brief. This is achieved with a two-stage LLM chain where one LLM extracts “ground truth” entities from the brief and a second LLM verifies their presence in the diagram.
The LLM, acting as a “Principal Architect”, performs a comprehensive critique of the generated architecture, evaluating both its analysis reports and diagrams.
The resulting critique provides ratings for \textit{Feasibility} and \textit{Clarity} on a $1$-$5$ scale (where $5$ is excellent), identifies key risks, and concludes with a recommendation for improvement. Finally, an assessment of the container diagram to identify potential vulnerabilities is carried out by an LLM with a “Cybersecurity Expert”, as part of the Security “Red Team” test.
This assessment provides an overall \textit{Risk Score}, calculated based on the number and severity of the identified vulnerabilities. In this case, a lower score indicates better security. As with any LLM-as-a-Judge method, these scores should be interpreted as heuristic indicators and may be affected by model bias, hallucinations, or domain miscalibration. In this study, we did not evaluate judge outputs against human expert ratings. Threats and mitigations are discussed in Section~\ref{sec:limits}.

\begin{figure*}[htbp]
    \centering
    \includegraphics[width=\linewidth]{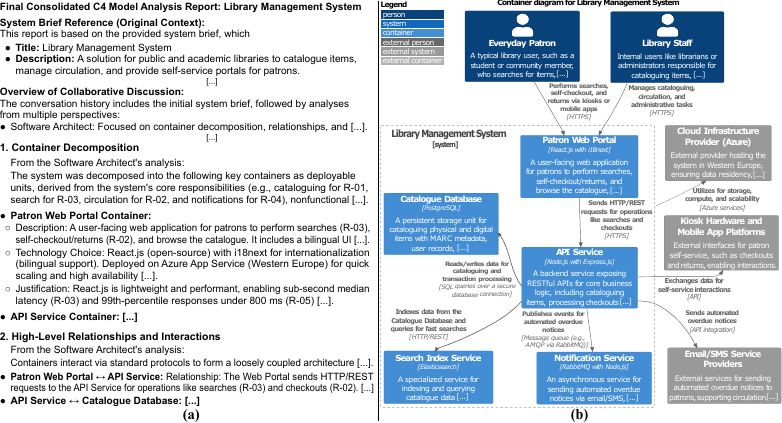}

\caption{Generated artifacts for the Library Management System test case: an excerpt from the \texttt{Analysis Report} (a) and its corresponding PlantUML diagram (b) at L2 (Container).  Artifacts were produced by the system using Grok~3~mini in the Collaborative Analysis (1 Round) configuration.}

    \label{fig:all_products}
\end{figure*}

\section{Experimental Setup}
\label{sec:experiments}

To evaluate our system, we ran experiments that varied two factors: (1) the underlying LLM and (2) the system configuration. This design lets us measure how each model and collaboration style affects C4 model quality across input briefs.

We used five diverse, canonical software system descriptions as input briefs, each representing common real‑world architectural challenges: \textit{Library Management System}, \textit{Next‑Generation Point‑of‑Sale System}, \textit{Online Bookstore}, \textit{Student Information System}, and \textit{Clinic Management System}. Each brief included a title, description, domain, constraints, and functional and non‑functional requirements to guide the agent‑based design process. Observing the qualitative results (system outputs) allows us to answer the first research question (RQ1).

To specifically address our second research question (RQ2) concerning the effectiveness of the collaborative workflow, we tested three distinct system configurations: (1)~Collaborative Analysis (3 Rounds), the whole, proposed system where persona-based agents engage in three rounds of conversation; (2)~Collaborative Analysis (1 Round), a streamlined version with only a single round of conversation to test the impact of conversational depth; and (3)~Single-Agent Analysis (Baseline), where a single agent replaces the collaborative step to provide a direct comparison against a non-collaborative approach.

We tested four distinct LLMs as the backend for the generation agents to understand how model choice affects architectural output quality: \texttt{gpt-4o}, \texttt{gpt-4o-mini}, \texttt{Grok 3 mini}, and \texttt{gemini-1.5-flash}. To ensure consistency and avoid bias in the qualitative assessments, a single model, \texttt{gemini-2.5-flash}, was used for all LLM-as-a-Judge evaluation tasks. In every case, the \texttt{temperature} parameter of LLMs was set to $0$ to ensure greater predictability of the system --- this, however, does not guarantee determinism of LLM outputs, and consequently the system's outputs.

\section{Results}
\label{sec:results}

\begin{table*}[t]
\centering
\caption{Summary of quantitative evaluation results. The table shows the performance of four LLM agents under three tested configurations based on the hybrid evaluation framework.}
\label{tab:model_performance_final_data}
\resizebox{\textwidth}{!}{%
\begin{tabular}{@{}llrrrrrrrr@{}}
\toprule

\textbf{Model} & \textbf{Configuration} &
\begin{tabular}[c]{@{}c@{}}Compilation\\ Success (\%)\end{tabular} &
Completeness (\%) & 
\begin{tabular}[c]{@{}c@{}}Abstraction\\ Adherence (\%)\end{tabular} &
\begin{tabular}[c]{@{}c@{}}Naming\\ Consistency (\%)\end{tabular} &
\begin{tabular}[c]{@{}c@{}}Semantic\\ Consistency (\%)\end{tabular} &

\begin{tabular}[c]{@{}c@{}}Architect\\ Clarity (1-5)\end{tabular} &
\begin{tabular}[c]{@{}c@{}}Architect\\ Feasibility (1-5)\end{tabular} &
\begin{tabular}[c]{@{}c@{}}Security\\ Risk Score (pts.)\end{tabular} \\
\midrule
\midrule

\multirow{3}{*}{GPT-4o} & Single-Agent & 80.11 & 100.00 & 97.50 & 60.85 & 51.71 & 4.8 & 4.0 & 30.2 \\
 & Collaborative (1 Round) & 89.78 & 100.00 & 100.00 & 16.46 & 25.62 & 2.4 & 2.8 & 31.6 \\
 & Collaborative (3 Rounds) & 100.00 & 100.00 & 100.00 & 8.24 & 34.12 & 3.2 & 3.6 & 36.6 \\
\midrule

\multirow{3}{*}{GPT-4o mini} & Single-Agent & 97.14 & 100.00 & 28.25 & 42.91 & 49.89 & 4.2 & 3.2 & 36.0 \\
 & Collaborative (1 Round) & 92.14 & 100.00 & 17.04 & 8.84 & 48.07 & 3.8 & 3.4 & 33.4 \\
 & Collaborative (3 Rounds) & 88.96 & 100.00 & 22.08 & 30.02 & 41.58 & 3.8 & 3.6 & 32.6 \\
\midrule

\multirow{3}{*}{Gemini 1.5 Flash} & Single-Agent & 66.19 & 100.00 & 78.86 & 80.42 & 48.68 & 4.0 & 3.4 & 33.2 \\
 & Collaborative (1 Round) & 100.00 & 81.46 & 95.00 & 82.00 & 29.52 & 3.0 & 2.4 & 34.0 \\
 & Collaborative (3 Rounds) & 95.00 & 77.10 & 78.33 & 66.52 & 30.64 & 2.8 & 1.8 & 36.5 \\
\midrule

\multirow{3}{*}{Grok 3 Mini} & Single-Agent & 58.10 & 100.00 & 100.00 & 61.36 & 48.98 & 4.8 & 3.8 & 36.8 \\
 & Collaborative (1 Round) & 72.14 & 100.00 & 100.00 & 24.74 & 43.82 & 3.6 & 3.4 & 30.6 \\
 & Collaborative (3 Rounds) & 73.21 & 98.33 & 100.00 & 14.14 & 41.86 & 3.4 & 3.2 & 27.4 \\

\bottomrule
\end{tabular}%
}
\end{table*}

This section presents the results of our experiments. First, we provide a qualitative review of the artifacts generated by the system for a representative test case to illustrate the nature and quality of the output. Second, we will present the quantitative results based on our evaluation framework across all models and configurations.

To demonstrate the system’s output, we present the exemplary artifacts for the “Library Management System” test case. Figure~\ref{fig: input} shows the full \texttt{System Brief}, the initial input to the collaborative analysis agents. This system description, including the title, domain, constraints, functional and non-functional requirements, formed the basis of the subsequent architectural reasoning of the agents.
From this input, at the container level (L2), the system generated the artifacts illustrated in Figure~\ref{fig:all_products}\,---\,Analysis Report and a corresponding PlantUML diagram. They present the characteristics of the produced outputs. The \texttt{Analysis Report} artifact is detailed and provides a strong narrative for architectural decisions. It has twelve headings: \textit{Container Decomposition}, \textit{High-Level Relationships and Interactions}, \textit{Implementation Feasibility}, and others. The diagram was generated based on the intermediate \texttt{View YAML} artifact, which encodes all architectural elements and relationships extracted from the \texttt{Analysis Report}. 
The diagram directly relates to the \texttt{Analysis Report} as it includes containers, technologies, and relationships mentioned in it.

The experimental results comparing performance across models and configurations are presented in Table~\ref{tab:model_performance_final_data}. Focusing first on the \textit{Structural \& Syntactic Integrity} metrics, we observe that\,---\,apart from the GPT-4o-mini model\,---\,\textit{Compilation Success} is higher for the collaborative configuration. \textit{Completeness} shows that most models generate all defined components; only Gemini 1.5 Flash performs poorly in this area, particularly under the collaborative setting. \textit{Abstraction Adherence} does not vary systematically with configuration type; however, Grok 3 mini and GPT-4o achieve the highest scores, indicating they produce the most accurate elements for the current C4 level. \textit{Naming Consistency} appears to depend on configuration, with the single-agent approach yielding the most consistent nomenclature.

Turning to the LLM-as-a-Judge metrics, \textit{Semantic Consistency} again favors the single-agent approach, which captures entities from the original system brief more accurately. Evaluations by the “Principal Architect” LLM-as-a-Judge model show that \textit{Clarity} and \textit{Feasibility} are also superior in the single-agent configuration. The \textit{Risk Score}, calculated as a weighted sum of points assigned by the “Risk Specialist” LLM-as-a-Judge model, is less informative as a standalone score; nevertheless, the accompanying list of potential vulnerabilities and their severity provides guidance for engineers working with the system.

As a reference, running the system in the configuration of the GPT-4o model in the 3-round collaborative setting consumed roughly 500{,}000 total tokens (input+output), whereas the single-agent baseline required about 40{,}000 tokens. For comparison, the system with GPT-4o-mini in the same 3-round setup used about 550{,}000 tokens, while its single-agent baseline was markedly cheaper at around 35{,}000 tokens.

Additionally, we can measure the complexity of the generated C4 model by counting the number of components identified at level L2. According to Table~\ref{tab:compact_component_stats}, the collaborative configuration produces more components overall. From a model perspective, the GPT models identify the largest number of components.


\section{Discussion}

This study set out to test whether a collaborative, LLM‑driven agent ensemble could rapidly generate a complete C4 model from a high‑level natural language brief, rather than to engineer the best‑performing end‑to‑end solution. The experiments confirm this feasibility: all three C4 abstraction levels were produced within minutes. For example, with a single‑round configuration using the Grok‑3‑Mini model, the end‑to‑end pipeline averaged roughly 10 minutes. However, runtime remains dependent on implementation, as the component level L3 could be performed in parallel. This illustrates the approach’s practical turnaround time.

A second matter worth mentioning is that the structural evaluation layer (PlantUML compilation and C4‑level compliance) proved reliable, while the semantic layer (LLM-as-a-Judge) delivered useful but occasionally noisy scores. Fine‑tuning the judge on C4‑specific vocabulary --- or equipping it with RAG (Retrieval Augmented Generation) --- should improve the meaningfulness of those metrics. Notably, the evaluation pipeline can be applied to diagrams drafted by human experts, providing engineers with an objective view through which to assess the quality of their own designs and to identify design gaps.

Throughout development, prompt engineering surfaced as a dominant factor: minor wording changes produced large shifts in agent output. Because the study aimed to observe collaborative dynamics rather than to exhaustively optimise prompts, some noise is inevitable. A more carefully engineered single‑agent baseline might therefore perform better than reported. Equally important is providing the agents with the right context. Agents that receive too little information about shared state and established elements of C4 struggle to maintain consistent terminology, especially since LLMs have context-window limits, and more context implies higher model call costs. Our observation is that even though the literature suggests the superiority of multi-agent systems, the orchestration and complexity of agents used in our experiments were not enough to prove that in our use case.

Results on naming consistency reveal that shared conversational memory is insufficient. Agents lack a common, persistent data structure on which to ground nomenclature, leading to divergent aliases for the same architectural element. Providing a shared glossary or a central knowledge graph may reduce this fragmentation.

The experiments do not clearly indicate which LLM is best suited for this system. GPT‑4o was the only non‑distilled model tested, yet it did not achieve higher metric scores. However, it identified the most components in both the Single‑Agent and C‑3 configurations (see Table~\ref{tab:compact_component_stats}), which may reflect that distilled LLMs contain a more compressed --- and therefore potentially less nuanced --- body of knowledge.

\begin{table}[t!]
\setlength{\tabcolsep}{4pt}      
\centering
\footnotesize                    
\begin{threeparttable}
  \caption{Component counts (mean, min, max) by model and configuration.}
  \label{tab:compact_component_stats}
  \begin{tabular}{llrrr}
    \toprule
    \textbf{Model} & \textbf{Config.\tnote{a}} &
      \multicolumn{3}{c}{\textbf{Components}} \\[-0.2em] 
    \cmidrule(lr){3-5}
    & & \textbf{Mean} & \textbf{Min} & \textbf{Max} \\
    \midrule
    \multirow{3}{*}{Gemini 1.5 Flash} & Single-Agent & 5.4 & 4 & 8 \\
                                      & C-1          & 1.8 & 0 & 4 \\
                                      & C-3          & 2.8 & 0 & 5 \\
    \midrule
    \multirow{3}{*}{Grok 3 Mini}      & Single-Agent & 4.6 & 4 & 5 \\
                                      & C-1          & 5.2 & 5 & 6 \\
                                      & C-3          & 5.4 & 5 & 6 \\
    \midrule
    \multirow{3}{*}{GPT-4o}           & Single-Agent & 7.0 & 6 & 8 \\
                                      & C-1          & 7.2 & 4 & 9 \\
                                      & C-3          & 8.2 & 7 & 9 \\
    \midrule
    \multirow{3}{*}{GPT-4o mini}      & Single-Agent       & 5.2 & 4 & 7 \\
                                      & C-1          & 7.6 & 6 & 9 \\
                                      & C-3          & 7.2 & 6 & 9 \\
    \bottomrule
  \end{tabular}
  \begin{tablenotes}[flushleft]
    \small
    \item[a] \textbf{Config.}: C-1 and C-3 denote collaborative configurations with 1 and 3 rounds, respectively.
  \end{tablenotes}
\end{threeparttable}
\setlength{\tabcolsep}{6pt}      
\end{table}

\section{Limitations and Future Work}
\label{sec:limits}

Our evaluation relies in part on an LLM-as-a-Judge procedure, which offers scalability and fast turnaround but introduces risks of bias, hallucinations, and limited domain expertise. We did not directly benchmark LLM judgments against human expert evaluations, which is a limitation. Future work will include a comparative human study to assess agreement and calibrate prompts, as well as robustness enhancements such as ensembling multiple LLM judges and augmenting judges with retrieval over literature on the C4 model.

Commercial deployments will require human-in-the-loop workflows in which engineers iteratively refine AI-generated artifacts; engineers should be able to inspect, accept, or correct each generated view while the system learns from those corrections. Further, richer memory mechanisms (e.g., vector stores indexed by design artifacts), cross-agent knowledge graphs, and dynamic prompt adaptation could unlock additional collaborative benefits in agentic LLM systems.

Incorporating the fourth level of the C4 model, Code, is a natural extension. It could be implemented analogously to the L3 component level and would require dedicated evaluation metrics, especially those covering code consistency and object-oriented design rules or other relevant methodologies.

Finally, the evaluation method itself can be improved by refining existing metrics and introducing new ones. For the LLM-as-a-Judge component, the clarity metric could be extended by inputting an image of the rendered diagram as LLMs’ ability to process graphical input develops.

\section{Conclusion}
\label{sec:conclusion_future_work}

\addtolength{\textheight}{-.2cm} 

This work presented an LLM‑driven multi‑agent platform that automates both the generation and evaluation of C4 architectural models. By emulating a collaboration of domain specialists, we showed that a concise, high‑level system brief can be transformed, within minutes, into a three‑tier C4 representation (Context, Container, and Component views).

In response to our research questions, we found that the multi-agent system is highly capable of generating C4 models (RQ1), though the quality varies with the underlying LLM. For instance, systems based on models like Grok 3 mini and GPT-4o achieved high abstraction adherence, while Gemini 1.5 Flash struggled with completeness. Regarding whether a multi-agent approach can outperform a single agent (RQ2), our results show that the single-agent baseline produced higher-quality artifacts in terms of semantic consistency, clarity, and feasibility. However, the multi-agent configuration consistently generated broader and more complex C4 models. This indicates that diverse and viewpoint-specific reasoning can bring additional architectural elements.

These results lead to three key takeaways. First, consistent with broader research on LLM-agentic systems, task decomposition combined with low-autonomy specialized agents produces reliable and repeatable results without excessive run time. Second, the evaluation method provides practical feedback on structural soundness, C4 compliance, semantic coverage, and security, giving practitioners an immediate diagnostic tool. Third, a simple multi-agent conversational collaboration for C4 level analysis is not superior to a single-agent configuration, and could require more advanced orchestration and agents.

In conclusion, while the resulting C4 artifacts are not yet production‑ready, they provide software architects with a rapid, information-rich starting point that is costly to obtain by hand. These findings are a meaningful step towards more automated software architecture design and may give valuable insight for researchers working on the development of LLM-based agentic systems.

\addtolength{\textheight}{-.2cm} 



\printbibliography

\end{document}